\documentclass[twocolumn]{NobArticle}

\title{Simulating MLB Seasons using Bayesian Inference and Random Walks}

\author{
    Simon Cha
}

\renewcommand{\maketitlehookd}{%
\begin{abstract}
    \noindent As a dedicated follower of sports statistics and with the MLB season beginning in late March, I set out to predict how many wins each team would accumulate by the end of the 162-game season. The goal was to build a simulation framework capable of forecasting the remainder of the season, starting from a 20-game burn-in period to establish initial estimates of team strength. My approach used a Bayesian inference model incorporating team win percentage, batting average, and pitching ERA to construct a posterior distribution of win probability for each matchup. For each game, I sampled from the posterior and simulated the outcome using a Bernoulli trial. Because future matchup inputs were unobserved, I forecasted batting averages using random walks and modeled pitching ERA with Kalman filters. After simulating many seasons, the model produced a distribution of win totals for all 30 teams and can also be used to estimate each team’s probability of making the postseason.
    
    \medskip
\end{abstract}
}
\usepackage{float}

\begin{document}

\small
\maketitle

\section{Introduction}
Forecasting long-term outcomes in sports presents a complex challenge due to evolving team dynamics, player performance variability, and incomplete future data. In this paper, I propose a simulation-based framework to forecast season win totals in Major League Baseball (MLB), using a Bayesian approach combined with synthetic performance projections. This methodology enables robust, forward-looking predictions even when full-season data is unavailable.
\section{Data Collection}
The Bayesian inference model in this project used three main input variables: team win percentage, team batting average, and starting pitcher ERA. Historical data was collected from the 2022, 2023, and 2024 MLB seasons. Because early-season batting and win percentage data are highly volatile due to small sample sizes, I filtered for games played between May 20th and August 20th. This mid-season window offers more stable metrics and reflects more reliable team and player performance levels. Pitching performance was measured using the starting pitcher's ERA on the day of each game, providing a more granular and context-specific metric compared to team-level pitching statistics. Data was scraped from several reputable sports analytics websites, including Baseball Reference, TeamRankings.com, SportsbookReview, and MLB.com. The Python libraries BeautifulSoup and Selenium were used to automate the data collection process.
\section{Methods}
The forecasting framework can be characterized as a three-step process. First, a Bayesian inference model is trained to predict the outcome of a game given input features. Second, synthetic inputs are generated for future games using statistical models. Finally, this process is repeated across many simulated seasons to produce a distribution of end-of-season win totals for each team.

\begin{figure}[!htpb]
    \centering
    \includegraphics[width=\linewidth]{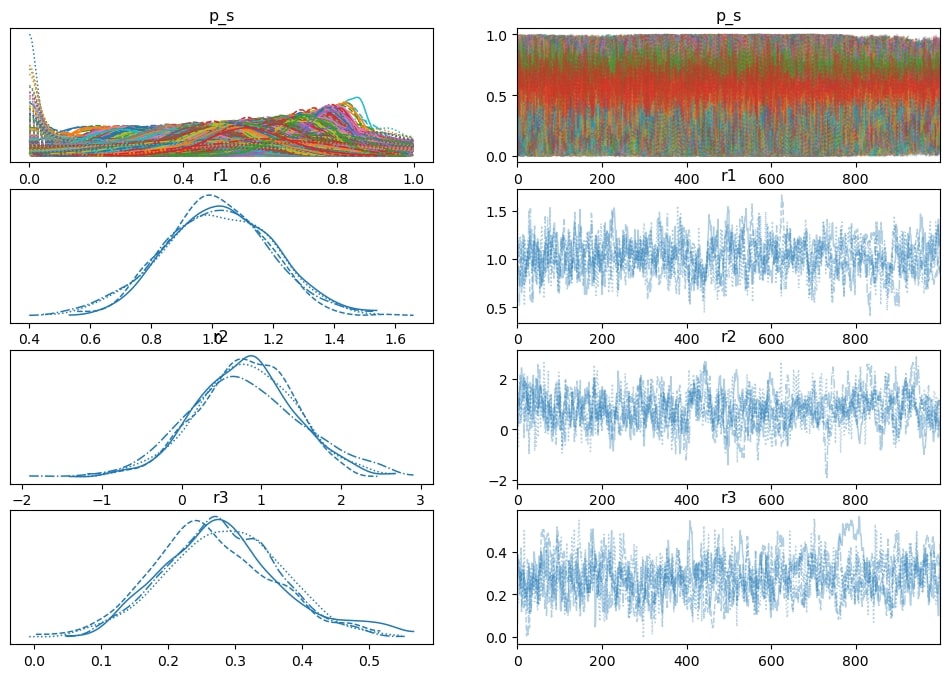}
    \caption{Posterior trace plots and density estimates for key model parameters: \( p_s \), \( r_1 \), \( r_2 \), and \( r_3 \). The right-hand column displays the MCMC trace plots showing sampled values across iterations, which are used to assess convergence and mixing. The left-hand column shows the marginal posterior distributions for each parameter.}
    \label{fig:tcanther}
\end{figure}
\subsection{Bayesian Inference}
This project builds upon the two-stage Bayesian modeling framework introduced by Yang and Swartz (2004), which modeled Major League Baseball (MLB) game outcomes using three key ratios: $\alpha$, $\beta$, and $\gamma$, representing win percentage, batting average, and starting pitcher ERA, respectively. These covariates were combined into a single relative strength metric through exponent-weighted contributions from each component, where the exponents $r_1$, $r_2$, and $r_3$ are contribution parameters learned from data and assigned independent uniform priors to reflect their relative importance.
\[
\lambda_s = \alpha_s^{r_1} \cdot \beta_s^{r_2} \cdot \gamma_s^{r_3}
\]
\noindent To account for uncertainty in the win probability, the model uses a two-stage Bayesian structure. First, the latent probability that the home team wins is drawn from a Beta distribution:
\[
p_s \sim \text{Beta}(m \cdot \lambda_s, m)
\]
\noindent where $m > 0$ controls the concentration of the distribution. Then, the game outcome is observed as a Bernoulli trial:
\[
X_s \sim \text{Bernoulli}(p_s)
\]
To estimate the posterior distributions of the contribution parameters \( r_1, r_2, r_3 \), I used Markov Chain Monte Carlo (MCMC) sampling. When the posterior distribution can't be computed directly, MCMC provides a practical way to approximate it by generating samples that reflect the distribution's shape. Similar to rejection sampling, it explores the parameter space by proposing values and accepting them based on how well they explain the data. This captures the uncertainty around each parameter and reveals how strongly each covariate contributes to the win probability model.

\subsection{Random Walks}
\noindent While forming posterior distributions of win probability for games with known pregame parameters is straightforward, forecasting future matchups—such as those in the 2025 season—poses a challenge, as the input parameters are not yet observed. To address this, I began by analyzing the evolution of team batting averages throughout the season. When normalized to reflect deviations from the mean (centered at zero), the trajectories interestingly resembled those of a Gaussian random walk: each new game introduced stochastic variation, with no strong directional trend over time.

\begin{figure}[!htpb]
    \centering
    \includegraphics[width=\linewidth]{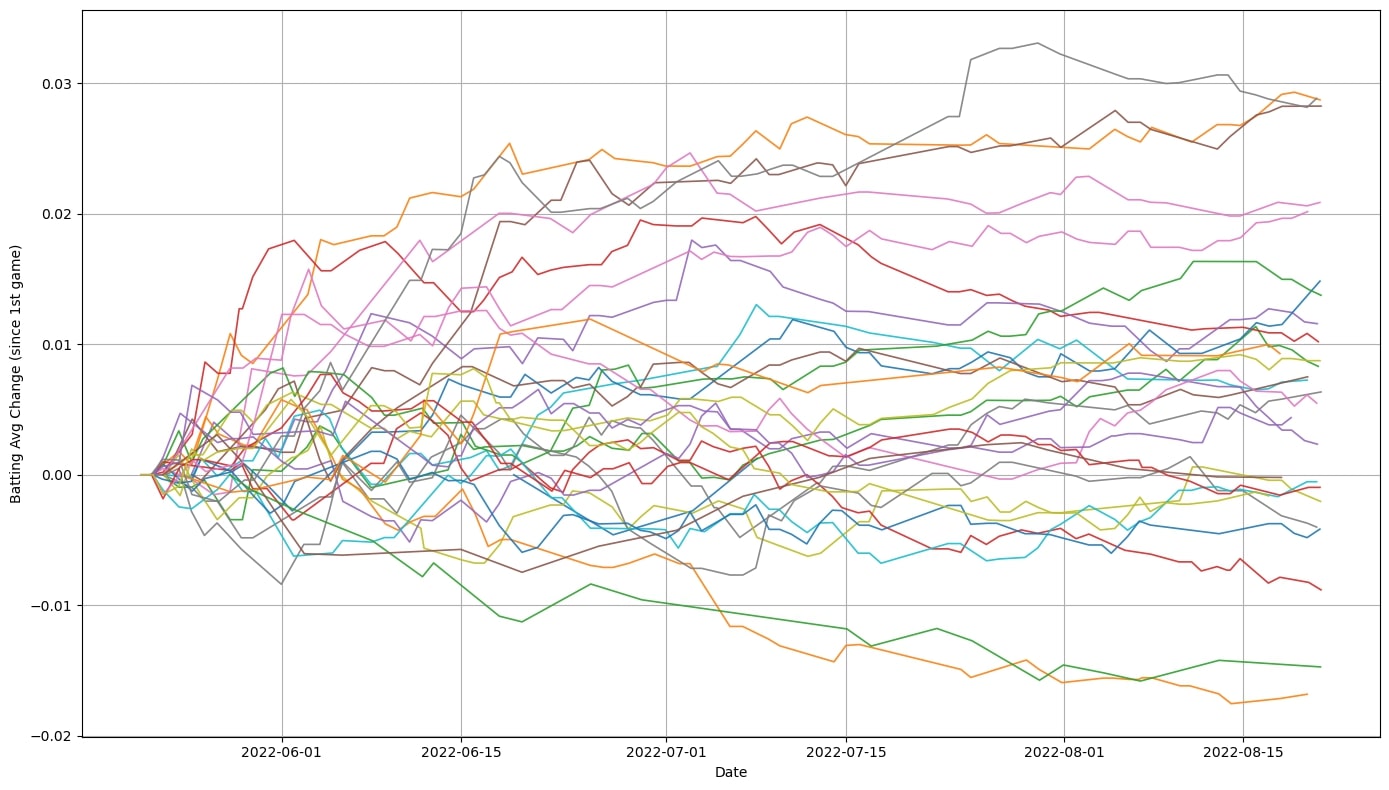}
    \caption{Batting averages of teams across the season modeled by a Gaussian Random Walk with Normal(0,0.0015).}
    \label{fig:tcanther}
\end{figure}

\subsection{Kalman Filters}

While batting average could be reasonably modeled as a Gaussian random walk, pitcher ERA presented a greater modeling challenge due to its high volatility and irregular jumps. To address this, I implemented a Kalman filter framework, which is better suited to separating true performance trends from noisy observations. The Kalman filter models ERA using a state-space formulation, where the hidden true pitching ability evolves slowly over time, and the observed ERA reflects that ability plus game-level noise. The model consists of two equations:
\begin{align*}
x_{t+1} &= x_t + w_t, \quad w_t \sim \mathcal{N}(0, \sigma^2_{\text{process}}) \\
y_t &= x_t + v_t, \quad v_t \sim \mathcal{N}(0, \sigma^2_{\text{obs}})
\end{align*}
Here, \( x_t \) represents the latent (true) ERA at time \( t \), and \( y_t \) is the observed ERA from a given game. The term \( w_t \) captures process noise—gradual drift in performance over time—while \( v_t \) captures observation noise, reflecting random variation from game to game. When forecasting future performance, we no longer have access to new observations, so the Kalman filter cannot perform its update step. Instead, predictions are generated by propagating the state forward using only the process model which is generally acceptable as the process noise is relatively small.

\begin{figure}[!htpb]
    \centering
    \includegraphics[width=\linewidth]{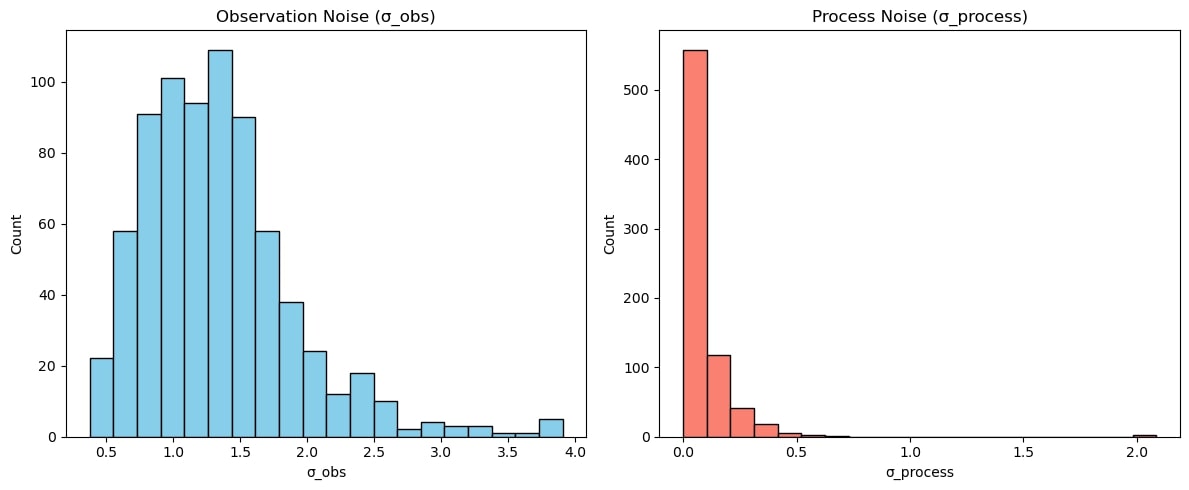}
    \caption{Distributions of estimated observation noise (\( \sigma_{\text{obs}} \)) and process noise (\( \sigma_{\text{process}} \)) from sliding window Kalman filter fits on team-level ERA data. The plots show that observation noise is generally much larger than process noise.}   
    \label{fig:tcanther}
\end{figure}
\noindent To estimate the observation and process noise components (\( \sigma_{\text{obs}} \) and \( \sigma_{\text{process}} \)) for the Kalman filter, I applied a sliding window approach using local level models from the \texttt{UnobservedComponents} class in \texttt{statsmodels}. For each team, 30-game windows of starting pitcher ERA data were fit with models that assume a slowly drifting latent performance level and noisy observed outcomes, producing rolling estimates of both noise terms. Teams were then grouped into terciles—low, medium, and high—based on their average ERA over the first 20 games of the season. To simulate future ERA trajectories, I sampled values of \( \sigma_{\text{obs}} \) and \( \sigma_{\text{process}} \) from the distributions observed within each group. These simulations reflect realistic ERA variation over time and serve as predictive inputs for the game-level Bayesian inference model.

\subsection{Simulation}
With synthetic input parameters for batting average and pitcher ERA generated for future games, we proceed to simulate full MLB seasons. Beginning after the designated burn-in period, we track each team's win percentage at the time of their first simulated game. This value is used as an input for upcoming matchups and is continuously updated as games are simulated through to the end of the regular season. For each game, we generate a posterior win probability using the Bayesian inference model and sample a single outcome from a Bernoulli distribution with that probability to determine the game result. By repeating this process 1{,}000 times, we generate a distribution of season win totals for each team.

\subsection{Limitations}
This paper has outlined the forecasting methodology but has not yet discussed its limitations and assumptions. A key concern is the simplification involved in modeling future game parameters. Batting average and pitcher ERA were treated as independent stochastic processes which ignores real-world dependencies like slumps, injuries, and lineup changes. While independence is assumed for tractability, some structural differences were accounted for: batting averages showed approximate exchangeability after normalization, whereas ERA required grouping teams into low, medium, and high tiers to reflect differing levels of variability. Another limitation is the use of data from Game 50 onward for model training. This choice was intentional to avoid the instability and high variance of early-season performance. While this improves reliability and generalizability, it limits the model's applicability to games earlier in the season or teams with little historical data.

\section{Results}
\begin{figure}[H]
\centering
\includegraphics[width=\linewidth]{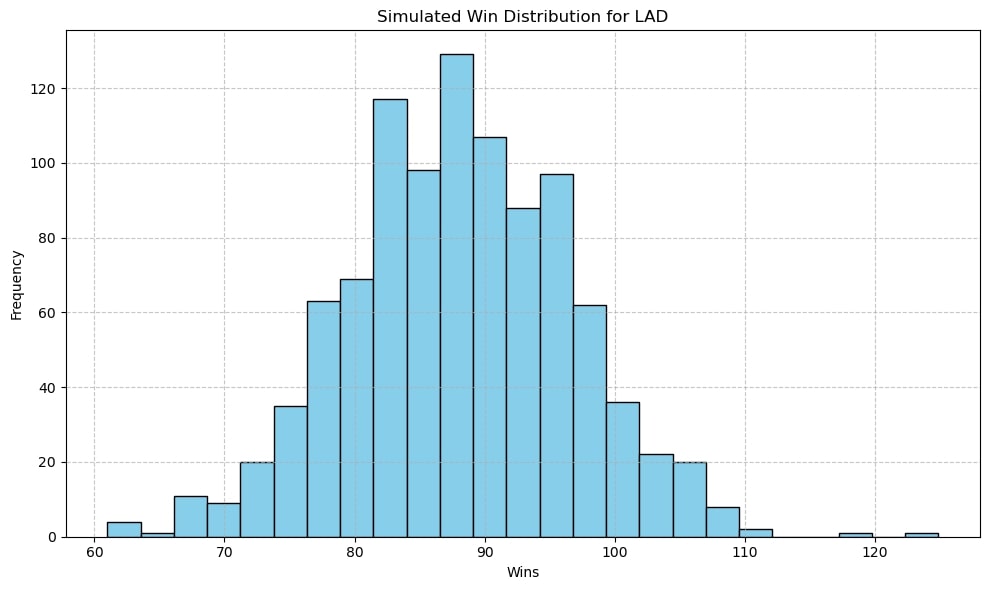}
\caption{Histogram of simulated win totals for the Los Angeles Dodgers (LAD) across 1,000 simulated MLB seasons, each consisting of 162 games.}
\label{fig:lad_win_hist}
\end{figure}

\begin{table}[H]
\centering
\begin{tabular}{lccc}
\toprule
\textbf{Team} & \textbf{Mean Wins} & \textbf{90\% CI (5, 95)} & \textbf{Playoff \%} \\
\midrule
SDP & 99.4 & (85.6, 112.4) & 90.9 \\
NYM & 96.1 & (80.5, 112.9) & 80.3 \\
PHI & 95.6 & (80.0, 112.8) & 78.6 \\
CHC & 93.3 & (78.0, 107.0) & 75.4 \\
DET & 93.1 & (78.4, 106.7) & 84.7 \\
TOR & 92.9 & (75.9, 110.5) & 79.5 \\
TBR & 90.3 & (72.4, 107.7) & 72.0 \\
SFG & 90.0 & (74.0, 104.9) & 62.2 \\
LAD & 88.1 & (74.0, 102.0) & 51.5 \\
NYY & 87.4 & (72.4, 101.6) & 62.6 \\
ARI & 85.5 & (70.0, 100.0) & 40.4 \\
TEX & 85.2 & (69.3, 100.6) & 52.4 \\
BOS & 83.6 & (67.8, 98.8) & 45.7 \\
HOU & 82.7 & (66.0, 98.8) & 43.1 \\
STL & 82.6 & (66.4, 98.0) & 31.7 \\
CIN & 82.5 & (65.0, 101.2) & 31.1 \\
LAA & 82.2 & (65.4, 98.0) & 42.9 \\
CLE & 82.1 & (65.8, 98.0) & 40.2 \\
MIL & 81.3 & (66.4, 95.8) & 26.3 \\
SEA & 79.9 & (63.8, 95.2) & 33.7 \\
ATH & 77.8 & (62.0, 93.0) & 0.0 \\
KCR & 77.7 & (59.8, 94.9) & 25.3 \\
WSN & 76.6 & (58.4, 95.5) & 15.8 \\
BAL & 72.9 & (56.0, 88.6) & 14.4 \\
MIA & 72.7 & (55.7, 89.1) & 7.3 \\
ATL & 71.5 & (53.3, 89.4) & 7.3 \\
PIT & 61.4 & (44.3, 79.4) & 1.0 \\
MIN & 60.7 & (42.3, 77.4) & 1.2 \\
CHW & 57.8 & (35.0, 79.4) & 2.3 \\
COL & 50.2 & (26.3, 73.4) & 0.2 \\
\bottomrule
\end{tabular}
\caption{Mean projected wins, 90\% confidence intervals, and playoff appearance probabilities across 1,000 simulated MLB seasons.}
\label{tab:win_playoff_summary}
\end{table}

\section{Discussion}
While I am not a baseball expert and cannot personally evaluate the realism of these rankings, I was encouraged to see meaningful variation in simulated win totals across teams. Interestingly, although the teams are sorted by descending mean wins, the playoff probabilities do not strictly follow this order. This is likely due to the nuanced structure of the league, where qualifying for the postseason depends not only on a team’s record but also on how it compares to others within the same division and league. However, the playoff probabilities appeared disproportionately extreme, with some teams exceeding a 90\% chance and others nearing 0\%. Given that the simulations begin after just 20 games, these estimates may be overly confident considering the inherent volatility of a 162-game season. This behavior likely reflects a combination of factors: teams with weak starts are heavily penalized, and the model inherently compounds these early trends as it extrapolates outcomes based on limited initial information.

\subsection{Future Directions}
For future directions, it would be valuable to incorporate additional features, as the current model is relatively simple and relies on just three variables: pitching ERA, batting average, and team win percentage. One potential enhancement is the integration of sports betting data as an additional prior, leveraging crowd-sourced probabilities to inform game outcomes. However, this approach presents challenges for long-term forecasting, since betting lines are typically only available a few days to a week in advance. Another promising direction involves modeling team momentum. Because baseball games are often played in multi-game series, the assumption of independence between consecutive games may overlook important dynamics. Facing the same opponent for multiple games in a row could introduce effects that are not captured under the current framework.

\section{Conclusion}

This project presented a probabilistic framework for forecasting MLB team performance using a combination of Bayesian inference, synthetic input generation, and season-long simulation. By modeling future batting and pitching parameters through Gaussian random walks and Kalman filters, the approach enabled full-season simulations even with limited early-season data. The resulting framework produced distributions of team win totals and playoff probabilities, offering a flexible method for exploring long-term outcomes under uncertainty. While the model relies on simplifying assumptions and limited features, it demonstrates how statistical tools can be used to approximate complex league dynamics. Future extensions could improve realism by incorporating additional covariates, leveraging betting markets, or modeling temporal dependencies such as momentum. Overall, this methodology provides a foundational step toward more nuanced simulation-based sports forecasting.

\section*{Acknowledgements}
I would like to acknowledge the foundational work of Yang and Swartz, whose two-stage Bayesian model for predicting Major League Baseball outcomes served as the conceptual basis for the modeling approach in this project.

\end{document}